\documentclass[twocolumn,showpacs,preprintnumbers]{revtex4}
\usepackage{amssymb}

\usepackage{graphicx}
\usepackage{dcolumn}
\usepackage{bm}

\input{tcilatex}

\begin{document}

\title{Noise spectroscopy of non-linear magneto optical resonances in Rb vapor. }
\author{M. Martinelli$^{2}$\thanks{Present address: Department of Physics, University 
of Toronto, Ontario, M5S
1A7 Canada}, P. Valente$^{1}$, H. Failache$^{1}$, D. Felinto$^{2}$, L. S. 
Cruz$^{2}$, P. Nussenzveig$^{2}$, and A. Lezama$^{1}$\thanks{E-mail: 
alezama@fing.edu.uy}}
\affiliation{$^{1}$ Instituto de F\'{i}sica, Facultad de Ingenier\'{i}a.
Casilla de correo 30, 11000, Montevideo, Uruguay} 
\affiliation{$^{2}$ Instituto de
F\'{i}sica, Universidade de S\~{a}o Paulo, Caixa Postal
66318, 05315-970, S\~{a}o Paulo, SP, Brazil }

\date{\today }

\begin{abstract}
Nonlinear magneto-optical (NMO) resonances occurring for near-zero magnetic
field are studied in Rb vapor using light-noise spectroscopy. With a
balanced detection polarimeter, we observe high contrast variations of the 
noise power (at fixed analysis frequency) carried by diode laser light 
resonant with the $5$S$_{1/2}\left(F=2\right) \rightarrow 5$P$_{1/2}\left( 
F=1\right) $ transition of $^{87}$Rb and transmitted through a rubidium vapor 
cell, as a function of magnetic field $B$. A symmetric resonance doublet
of anti-correlated noise is observed for orthogonal polarizations around 
$B=0 $ as a manifestation of ground state coherence. We also observe sideband 
noise resonances when the magnetic field produces an atomic Larmor
precession at a frequency corresponding to one half of the analysis
frequency. The resonances on the light fluctuations are the consequence of
phase to amplitude noise conversion owing to nonlinear coherence effects in
the response of the atomic medium to the fluctuating field. A theoretical
model (derived from linearized Bloch equations) is presented that reproduces
the main qualitative features of the experimental signals under simple
assumptions.
\end{abstract}


\maketitle

\preprint{}

\section{Introduction.}

\subsection{Noise spectroscopy.}

Light waves such as laser beams possess unavoidable fluctuations either in
amplitude, frequency or phase. Such fluctuations may arise from the specific
dynamics of the light source and are, in any case, submitted to the
constraints imposed by the quantum nature of light. The light field
fluctuations determine its spectrum. In most cases, the fluctuations in the
light wave play a significant role in the interaction with matter. A trivial
example is the broadening of spectral features as a result of the finite
spectral width of the light \cite{GEORGES79}. The study of the detailed
influence of light fluctuations in the light-atom interaction has motivated
a large amount of research \cite{GEORGES79,AGARWAL78,GEORGES80,DALTON82,
DALTON82B,ELLIOTT85,ANDERSON90,RITSCH90,VEMURI91,CAMPARO91,JYOTSNA95,KINROT95,CAMPARO93}%
. In some cases, the light fluctuations are modelled as classical stochastic
processes \cite{AGARWAL78}, in other cases a fully quantum approach is
required \cite{DALTON82,DALTON82B}. It has been demonstrated that several
features of the atomic response are significantly affected by the light
fluctuations. In turn, the interaction with the atomic system may result in
interesting modifications of the light field statistics giving rise to such
effects as anti-bunching, squeezing, etc. \cite{SCULLYBOOK97}.

A renewal of interest in the study of light fluctuations in matter-wave
interactions occurred after Yabusaki \textit{et al.}~\cite{YABUSAKI91}
demonstrated that the study of fluctuations of the light transmitted through
an atomic sample can be used as a spectroscopic tool~\cite
{MCINTYRE93,WALSER94,ROSENBLUH98,CAMPARO99}. Indeed, the spectral analysis
of the photocurrent produced in a detector by the light transmitted through
the sample reveals spectral features that are associated to the atomic
dynamics. In the experiment of~\cite{YABUSAKI91} a diode laser was used.
Diode laser light is known to possess (under rather ordinary conditions)
small amplitude fluctuations and large phase fluctuations \cite{GIACOBINO95}%
. If the laser beam is directly sent on a detector (such as a photodiode)
which is sensitive only to light intensity, a low level of fluctuations is
observed (essentially shot-noise). On the other hand, if the light beam
reaches the detector \emph{after} traversing a resonant atomic medium a
substantial increase in the fluctuation level is observed. One can say that
the atomic medium is responsible for transforming the phase noise (present
in the incident field) into amplitude noise that can be observed with the
photodetector \cite{WALSER94,CAMPARO99}.

The simplest way for achieving the transformation from phase noise into
intensity noise is to produce interference of the field with a second
(reference) field of same frequency (homodyning) which is out of phase. Let
the first field be $\alpha \left( t\right) =\left[ \bar{\alpha}+\delta
\alpha \left( t\right) \right] e^{i\omega t}+cc$ where $\bar{\alpha}$ is a
real constant amplitude and the fluctuation term $\delta \alpha \left(
t\right) =ib\left( t\right) $ is purely imaginary (\textit{i.e.} represents
phase noise if $\left| b\left( t\right) \right| \ll \bar{\alpha}$). Let $%
\beta \left( t\right) =\left[ A+iB\right] e^{i\omega t}+cc$ be a
monochromatic reference field with complex amplitude and negligible
fluctuations. Then the fluctuation in the total field intensity is given by: 
\begin{equation}
\delta I\left( t\right) =2Bb\left( t\right) \;,  \label{delta I}
\end{equation}
indicating that the intensity noise is proportional to the incident field
phase noise and the quadrature component of the reference field.

For optically thin atomic samples, one can describe \cite{WALSER94} the
transmitted field $E_{T}$ as: 
\begin{equation}
E_{T}\left( t\right) =\left[ \mathbf{\alpha }\left( t\right) +i\mathbf{\beta 
}\left( t\right) \right] e^{i\omega t}\;,  \label{campo total en el detector}
\end{equation}
where $\mathbf{\alpha }\left( t\right) $ is the incident field and $\mathbf{%
\beta }\left( t\right) $ is a field radiated by the atoms that is
proportional to the atomic electric dipole. Consequently, the field
generated by the atoms plays the role of the reference field in the homodyne
detection. From Eqs. \ref{delta I} and \ref{campo total en el detector} we
see that the phase fluctuations present in the incident field will be
transformed into light intensity fluctuations by the real part of the atomic
dipole which is directly linked to the dispersive properties of the medium.
It is clear from this analysis that the observation of the transmitted field
noise, as a function of different parameters, provides an useful handle to
study the influence of such parameters on the atomic dispersion. In the
preceding discussion we have neglected fluctuations in the atomic dipole
responding to the fluctuations in the incident field. If such atomic
fluctuations are included in the analysis they result in an additional
contribution to the transmitted field noise carrying information on the
atomic dynamics. In particular, one should expect a resonant behavior of the
transmitted noise whenever a spectral component of the incident field
fluctuation corresponds to a characteristic frequency of the atom-field
evolution.

\subsection{Nonlinear magneto-optical effects.}

When a light field resonantly interacts with an atomic transition between
two levels with angular momentum degeneracy, the different polarization
components of the incident field couple to different transitions between
Zeeman sublevels. As a result, the absorption and dispersion experienced by
the different polarization components may be different, giving rise to such
effects as birefringence and dichroism which result in changes in the
polarization (as well as amplitude and phase) of the transmitted light.
These effects depend on the specific atomic Zeeman sublevel energies and are
consequently very sensitive to the presence of an external magnetic field.
Such magneto-optical effects have been recently reviewed by Budker \textit{%
et al.}~\cite{BUDKER02}. In this comprehensive review a distinction is
established between linear magneto-optical effects (depending linearly on
the light field intensity) and nonlinear magneto-optical effects (NMOE)
involving the quadratic or higher order dependence on light intensity. While
linear magneto-optics is a quite old branch of optical physics, NMOE has
attracted considerable attention in recent years. Such attention was
motivated by the discovery in experiments on atomic transitions with a
long-lived lower state, of a very large enhancement in the magneto-optical
response of the atomic sample when the optical field verifies the condition
for a two-photon (Raman) resonance between ground state Zeeman sublevels. In
the case of a single optical field interacting with the atoms, the Raman
resonance condition for two orthogonal polarization components occurs at
zero magnetic field. The enhanced magneto-optical response is nonlinear in
the light intensity and is due to the coherence induced by the field between
ground state Zeeman sublevels. A typical example of such NMOE is the
nonlinear Faraday effect\ resulting in a steep variation of the optical
field polarization angle when a longitudinal magnetic field is tuned across $%
B=0$. A distinctive feature of such NMOE is its narrow width, which
corresponds, in terms of magnetic field, to $g\mu _{B}B\sim \hbar \gamma $,
where $\gamma $ is the relaxation rate acting upon the ground state Zeeman
coherence, $g$ is the Land\'{e} factor and $\mu _{B}$ the Bohr magneton.
Since ground state dephasing collisions are usually negligible in actual
experimental conditions involving alkali atoms, $\gamma $ is essentially
determined by the interaction time between light and atom. Several
techniques can be used to obtain long (effective) interaction times such as
Ramsey zones, paraffin coated cells or vapor cells with buffer gas. NMOE
resonances as narrow as $2\pi \times 1$ Hz have been observed \cite{BUDKER98}%
. The nonlinear Faraday effect is intimately related to other NMOE such as
ground state Hanle effect, Hanle/CPT resonances, self rotation, alignment to
orientation conversion, slow group velocity, etc. that are the consequence
of the coherence created in the atomic ground state \cite{BUDKER02}.

The experimental observation of NMOE imposes some demanding experimental
conditions. A careful shielding and control of the magnetic field is
necessary. Also, in order to detect the small NMOE from background noise and
systematic deviations, some kind of modulation technique is normally
required. The nonlinear Faraday effect has been investigated using high
precision polarimetry in which the incident field polarization was modulated
by a Faraday rotator \cite{BUDKER98,BUDKER99}. In the balanced detection
scheme, the light transmitted through the sample is separated into two
branches with orthogonal polarizations and balanced intensities. As the
magnetic field is scanned, the difference between the intensities of the two
branches is monitored. A lock-in amplifier extracts the signal at the first
or second harmonic of the modulation frequency. Recently, frequency
modulation of light was used for the detection of nonlinear magneto-optical
rotation \cite{BUDKER02B}. In such a case, in addition to the usual
nonlinear magneto-optical (NMO) resonance occurring for $B\simeq 0$,
symmetric sidebands occur when the magnetic field gives rise to a Larmor
precession frequency equal to a multiple of the FM frequency (that can be
easily achieved in the MHz range). These sidebands whose intrinsic width is,
as for the central feature, determined by $\gamma $ allow the extension of
NMOE based magnetometry to fields of the order of one Gauss. Recently,
narrow NMOE resonances have been observed in Rb vapor in the presence of Ne
buffer gas \cite{FAILACHE03} using a parametric resonance technique \cite
{DUPONT70}. In this case the magnetic field is modulated. As for the FM NMOE
spectroscopy, sharp sidebands are observed when the modulation frequency
equals twice the atomic Larmor precession in the magnetic field.

The purpose of this paper is to demonstrate that noise spectroscopy can be
used as a convenient sensitive tool for the observation of nonlinear
magneto-optical effects. The method consists of taking advantage of the
intrinsic phase modulation of the light field produced by its random
fluctuations. FM and noise spectroscopy are closely related techniques as
noticed in \cite{MCINTYRE93,ROSENBLUH98}. In both cases the atomic medium
introduces an imbalance between symmetric spectral field components
resulting in an intensity modulation of the light.

The observation of noise resonances associated to the transient precession
(triggered by spontaneous emission events) of oriented atoms in a magnetic
field was recently reported \cite{MITSUI00}. In addition, the connection
between light fluctuation statistics and NMOE has come into focus after the
proposal \cite{MATSKO02} and the recent experimental observation \cite
{RIES03,JOSSE03} of vacuum squeezing via self rotation.

A close connection exists between NMOE and other coherence phenomena such as
electromagnetically induced transparency (EIT). EIT is frequently realized
in three level $\Lambda $ systems where the two lower levels are Zeeman
sublevels of the same degenerate atomic level and the fields acting on the
two arms of the $\Lambda $ system are the two circular components of a
linearly polarized optical wave \cite{AKULSHIN98,PHILLIPS01}. Notice that,
in this case, a perfect (classical) correlation exists between the
fluctuations on the two fields participating in \ the EIT. The resonance
between the fields and the atomic levels is controlled by a longitudinal
magnetic field. From this point of view, the zero-field NMO resonance
correspond to the Raman resonance condition and thus to the occurrence of
EIT owing to coherent population trapping (CPT) in the ground state. Due to
its potential application to quantum optical information processing~\cite
{LUKIN01}, there is currently a large interest in the statistical properties
of \ light fields participating in EIT resonances~\cite{GARRIDO03}. From
this perspective, the following study of the noise properties of the
transmitted light by an atomic sample in the vicinity of a NMO resonance
also concerns the statistical properties of two perfectly correlated
classical fields undergoing an EIT resonance.

\section{Experiment.}

\begin{figure}[tbp]
\includegraphics[width=3.5in]{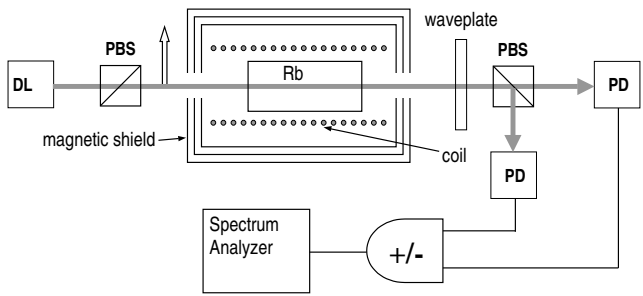}
\caption{Experimental setup. DL: diode laser; PBS: polarizing beam splitter;
PD: photodiode.}
\label{Setup}
\end{figure}

The experimental setup is sketched in Fig. \ref{Setup}. It consisted of a
diode laser, frequency stabilized (with a saturated-absorption-based
control) to the $^{87}$Rb D1 line $F_{g}=2$ to $F_{e}=1$ transition. The
laser output was linearly polarized and sent through a 5 cm long Rb vapor
cell (room temperature, no buffer gas). The total light power available at
the cell was 3 mW (0.5 cm$^{2}$ beam cross section). The cell was placed
inside a three-layer mu-metal magnetic shield surrounding a solenoid that
allows the scanning of a longitudinal magnetic field. After the cell, the
light polarization was analyzed with a balanced polarimeter consisting of a
waveplate and a polarizing beam splitter cube. Two different polarization
decompositions of the light were used. When a $\lambda /4$ waveplate was
present between the cell and the beam splitter the light was separated into
the two orthogonal circular components ($\sigma ^{+}$ and $\sigma ^{-}$).
If, instead, a $\lambda /2$ waveplate was used, the light was decomposed
into two linear and orthogonal polarizations at $\pm 45^{\circ }$ with
respect to the incident field polarization. The light intensities along the
two output branches of the polarimeter were detected by similar (balanced)
photodiodes with 10 MHz bandwidth. The two photocurrents were added or
subtracted before being sent to an electronic spectrum analyzer with 1.8 GHz
bandwidth. The spectrum analyzer was used in the zero span mode thus
operating as a band-pass filter. A 2.5 MHz analysis frequency (30 kHz
resolution bandwidth) was used since this frequency happened to be well
separated from spurious ambient noise. The noise power was recorded as a
function of the magnetic field at the cell. The experiments were carried
with an extended cavity diode laser (linewidth $\lesssim $ 1 MHz) whose
field fluctuations are spectrally narrower if compared with free running
lasers. This choice was due to practical considerations related to the
improved frequency stability and tunability of extended cavity diode lasers.
We have carried some preliminary tests indicating that the use of a free
running laser is also possible. Since in that case the field fluctuations
have a broader spectrum, one may expect that the use of a free running laser
will favor the utilization of larger analysis frequencies \cite{CAMPARO99}.

\begin{figure}[tbp]
\includegraphics[width=3.5in]{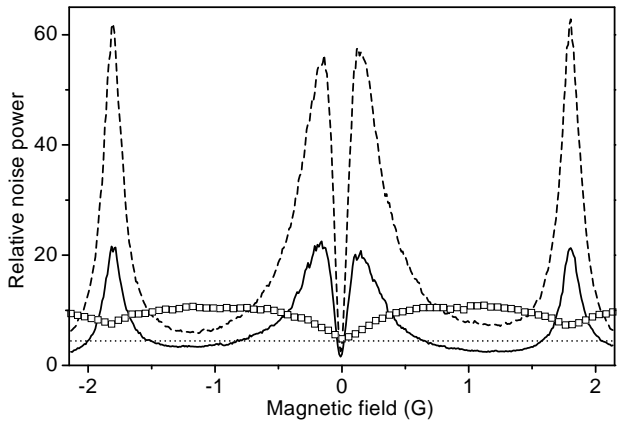}
\caption{Observed noise power relative to shot-noise at 2.5 MHz as a
function of magnetic field. Solid line: photocurrent difference noise for
circular polarizations. Dashed line: photocurrent difference noise for
linear polarizations. Hollow squares: Noise on the photocurrent sum for both
polarization choices. Dotted line: off-resonance noise background on
photocurrent sum. }
\label{Signal}
\end{figure}

The recorded signals corresponding to the noise power on the sum and
difference of the two photocurrents as a function of magnetic field, for two
different choices of the polarization components of the polarimeter, are
presented in Fig.~\ref{Signal}. The noise power is presented relative to the
shot-noise level measured on the photocurrent difference when the laser
frequency is tuned away from the atomic resonance. Also shown by the dotted
line in Fig. \ref{Signal} is the excess intensity-noise level (6.5 $\pm $
0.1 dB above the shot-noise) existing on the off-resonance light. Around $B=0
$, the noise on the photocurrent sum presents a reduction (to a level
approaching the excess-noise level) due to the occurrence of EIT. In
addition, smaller reduction of the sum noise are observed on two sidebands
occurring when the magnetic field induces a Zeeman shift of the ground state
Zeeman sublevels corresponding to one half of the analysis frequency (The
Zeeman shift for the 5P$_{1/2}\left( F=2\right) $ level of $^{87}$Rb is $0.7$
MHz/Gauss). The photocurrent difference noise is dominant for small values
of $B$ (corresponding to the occurrence of the EIT resonance) and at the
sidebands indicating anti-correlation of the field fluctuations. A sharp ($%
100\%$ contrast) decrease in the photocurrent difference noise occurs for
exact cancellation of the magnetic field where the noise drops below the
off-resonance excess-noise level (dotted line) down to the shot-noise level
(corresponding to 1 in the units used on the vertical axis of Fig. \ref
{Signal}). As discussed below, this complete cancellation of the atomic
contribution to the photocurrent difference noise is a consequence of the
symmetry existing between the two polarization components when $B=0$. Rather
similar shapes in the magnetic field dependence of the photocurrent
difference noise are observed for circular and linear polarizations. In the
latter case the noise power is increased ($4.1$ $\pm 0.7$ dB around $B=0$).
The intensity dependence of the central feature of the photocurrent
difference noise for circular polarizations is shown in Fig. \ref{Power
dependence}. The doublet narrows and decreases as the light intensity is
reduced. The narrowest observed doublet corresponds to a 70 mG separation
between maxima. A similar narrowing and reduction is observed on the central
dip on the photodetector sum signal noise (not shown).

\begin{figure}[tbp]
\includegraphics[width=3.5in]{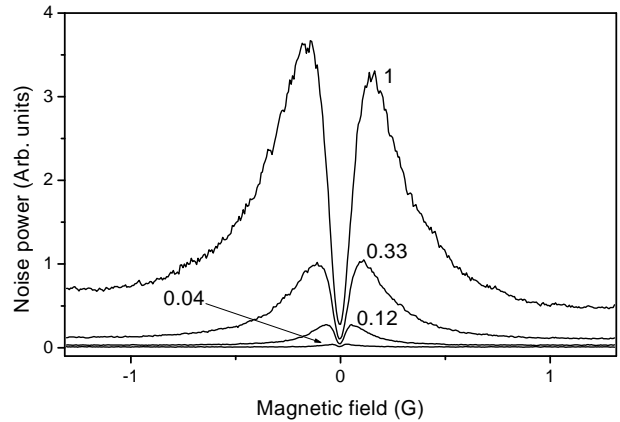}
\caption{Central structure of the noise power signal at 2.5 MHz for circular
polarizations as a function of magnetic field for various light intensities.
The relative light power is indicated. }
\label{Power dependence}
\end{figure}

\section{Theory.}

We present in this section a simple theoretical analysis of the interaction
of an atomic system and a fluctuating light field leading to the calculation
of balanced detection signals. The model is intended for an
electric-dipole-allowed atomic transition between two energy levels with
arbitrary angular momentum degeneracy. The transmitted field is analyzed
along arbitrary polarization components. The calculation is a direct
extension of previous treatments \cite{LEZAMA00,VALENTE03} with the
inclusion of a fluctuating component on the incident field. The field is
assumed to present small time dependent fluctuations around a constant
nonzero mean value. This assumption allows the calculation of the atomic
fluctuations through a linear response theory \cite{JYOTSNA95}.

We consider two degenerate levels: a ground level $g$ of total angular
momentum $F_{g}$ and energy $0$ and an excited state $e$ of angular momentum 
$F_{e}$ and energy $\hbar \omega _{0}$. The total radiative relaxation
coefficient of level $e$ is $\Gamma $. We assume that the atoms in the
excited state radiatively decay into the ground state $g\ $at the rate $\eta
\Gamma $, where $\eta $ is a branching ratio coefficient that depends on the
specific atomic transition [$0\leq \eta \leq 1$; $\eta =1$ for a closed
(cycling) transition]. We consider a homogeneous ensemble of atoms at rest.
However, in order to simulate the effect of a finite interaction time of the
atoms with the light, we assume that the atoms escape the interaction region
at a rate $\gamma $ ($\gamma \ll \Gamma $). This escape is compensated, in
steady-state, by the arrival of fresh atoms in the ground state.

The incident field is $\mathbf{E}(t)=E\mathbf{e}e^{i\omega _Lt}+E^{*}\mathbf{%
e}^{*}e^{-i\omega _Lt}$ where $E$ is the complex amplitude and $\mathbf{e}$
is a unit polarization vector.

The atom field coupling in the rotating wave approximation (RWA) is: 
\begin{eqnarray}
W &=&\hbar \Omega \mathbf{e}.\mathbf{Q}_{ge}e^{i\omega _Lt}+HC
\label{interaccion campo atomo} \\
&=&\hbar \left( V_{ge}e^{i\omega _Lt}+V_{eg}e^{-i\omega _Lt}\right) \;. 
\nonumber
\end{eqnarray}

\noindent Here $\mathbf{Q}_{ge}=P_{g}\mathbf{Q}P_{e}$ where $P_{g}$ and $%
P_{e}$ are the projectors on ground and excited state manifolds
respectively. The dimensionless operator $\mathbf{Q}$ is related to the
atomic electric dipole operator through: 
\begin{equation}
\mathbf{D}=\left\langle g\Vert \mathbf{D}\Vert e\right\rangle \mathbf{Q} \;,
\label{definicion operador Q}
\end{equation}
where $\left\langle g\Vert \mathbf{D}\Vert e\right\rangle $ is the reduced
matrix element of the dipole operator between the ground and excited state
(considered real). $2\Omega =\left\langle g\Vert \mathbf{D}\Vert
e\right\rangle E/\hbar $ is the reduced Rabi frequency of the field.

The total Hamiltonian in the RWA is: $H=H_{0}+W$ where $H_{0}=\hbar \omega
_{0}P_{e}+(\beta _{g}P_{g}+\beta _{e}P_{e})F_{z}B$ is the atomic Hamiltonian
including the Zeeman coupling with the magnetic field $B$ ($\beta _{g}$ and $%
\beta _{e}$ are the ground and excited sate gyromagnetic factors and $F_{z}$
is the total angular momentum projection along the magnetic field).

The evolution of the atomic density matrix $\rho $ is governed by the Bloch
equation: 
\begin{eqnarray}
\dot{\rho} &=&-\frac{i}{\hbar }\left[ H_{0}+W,\rho \right] -\frac{\Gamma }{2}%
\left\{ P_{e},\rho \right\}  \label{bloch} \\
&&+\eta \Gamma \left( 2F_{e}+1\right) \sum_{q=-1,0,1}Q_{ge}^{q}\rho
Q_{eg}^{q}-\gamma \left( \rho -\rho _{0}\right) \;.  \nonumber
\end{eqnarray}

$\ Q_{ge}^q=Q_{eg}^{q\dagger }\ \ (q=-1,0,1)$ are the standard components of 
$\mathbf{Q}$ and $\gamma \rho _0$ represents a constant pumping rate (due to
the arrival of fresh atoms) in the isotropic state $\rho _0=P_g/\left(
2F_g+1\right) $.

After traversing an optically thin atomic medium, the transmitted field is
given by \cite{WALSER94}: 
\begin{eqnarray}
\mathbf{\varepsilon }\left( t\right) &=&\left( \mathbf{E}+i\kappa \mathbf{P}%
\right) e^{i\omega _{L}t}  \label{campo con polarizacion q} \\
&=&\left[ \mathbf{\alpha }\left( t\right) +i\mathbf{\beta }\left( t\right) %
\right] e^{i\omega _{L}t}\;,  \label{campo con polarizacion q simplificado}
\end{eqnarray}
where $\kappa $ is a constant dependent on the optical thickness of the
sample. We assume that both $\mathbf{\alpha }\left( t\right) $ and $\mathbf{%
\beta }\left( t\right) $ are stationary fluctuating (complex) quantities
corresponding to the incident field and to the field radiated by the atoms
respectively. We write: 
\begin{eqnarray}
\mathbf{\alpha }\left( t\right) &=&\left[ \bar{\alpha}+\delta \alpha \left(
t\right) \right] \mathbf{e}  \label{campo fluctuante} \\
\mathbf{\beta }\left( t\right) &=&\mathbf{\bar{\beta}}+\delta \mathbf{\beta }%
\left( t\right) \;,  \label{dipolo fluctuante}
\end{eqnarray}
with $\overline{\delta \alpha \left( t\right) }=0$ and $\overline{\delta 
\mathbf{\beta }\left( t\right) }=0$ (the upper bar indicating a stochastic
average).

The output field component with complex polarization $\mathbf{q}$ is $%
\varepsilon _{q}\left( t\right) \equiv \mathbf{q}^{\ast }\cdot \mathbf{%
\varepsilon }\left( t\right) \equiv \bar{\varepsilon}_{q}+\delta \varepsilon
_{q}\left( t\right) \equiv \left( \alpha _{q}+i\beta _{q}\right) +\left[
\delta \alpha _{q}\left( t\right) +i\delta \beta _{q}\left( t\right) \right] 
$ and the corresponding intensity: 
\begin{eqnarray}
I_{q}\left( t\right) &=&\left| \bar{\varepsilon}_{q}\right| ^{2}+\delta
I_{q}\left( t\right)  \label{intensidad en funcion de t} \\
\delta I_{q}\left( t\right) &\simeq &\left( \alpha _{q}^{\ast }-i\beta
_{q}^{\ast }\right) \left[ \delta \alpha _{q}\left( t\right) +i\delta \beta
_{q}\left( t\right) \right] +cc \;,  \label{fluctuacion de intensidad}
\end{eqnarray}
where only linear noise contributions are retained.

Taking the Fourier transform of Eq. \ref{fluctuacion de intensidad} we get: 
\begin{eqnarray}
\widetilde{\delta I}_q\left( \omega \right) &\simeq &\widetilde{\delta I}%
_q^F\left( \omega \right) +\widetilde{\delta I}_q^A\left( \omega \right)
\label{delta I de omega} \\
\widetilde{\delta I}_q^F\left( \omega \right) &=&\left( \alpha _q^{*}-i\beta
_q^{*}\right) \widetilde{\delta \alpha }_q\left( \omega \right)  \nonumber \\
&&+\left( \alpha _q+i\beta _q\right) \widetilde{\delta \alpha }_q^{*}\left(
-\omega \right)  \label{delta I field} \\
\widetilde{\delta I}_q^A\left( \omega \right) &=&i\left( \alpha
_q^{*}-i\beta _q^{*}\right) \widetilde{\delta \beta }_q\left( \omega \right)
\nonumber \\
&&-i\left( \alpha _q+i\beta _q\right) \widetilde{\delta \beta }_q^{*}\left(
-\omega \right) \;,  \label{delta I atomo}
\end{eqnarray}
where we have identified two contributions to the light intensity
fluctuations: $\widetilde{\delta I}_q^F\left( \omega \right) $ arising from
the incident field fluctuations and $\widetilde{\delta I}_q^A\left( \omega
\right) $ originating from the induced atomic dipole fluctuations. If the
atomic dipole fluctuations can be neglected, then $\widetilde{\delta I}%
_q\left( \omega \right) \simeq \widetilde{\delta I}_q^F\left( \omega \right) 
$ and the only terms producing phase to amplitude noise conversion are
proportional to the real part of the mean atomic dipole moment as was
assumed in Eq. \ref{delta I}.

In general, the atomic dipole fluctuations are not negligible. We evaluate
their contribution using a linear response approach. For this, we write the
incident field fluctuations in the form $\delta \alpha \left( t\right)
=a\left( t\right) +ib\left( t\right) $ were $a\left( t\right) $ and $b\left(
t\right) $ are real fluctuating functions with zero average value. Since $%
\bar{\alpha}$ has been taken real, $a\left( t\right) $ [Fourier transform: $%
\widetilde{a}\left( \omega \right) $] represents in-phase (amplitude)
fluctuations while $b\left( t\right) $ [Fourier transform: $\widetilde{b}%
\left( \omega \right) $] corresponds to quadrature fluctuations. In the
limit of small fluctuations $b\left( t\right) $ represent phase noise. The
calculation of the spectrum of transmitted field intensity fluctuations, $%
\widetilde{\delta I}_{q}\left( \omega \right) $, as a function of $\ 
\widetilde{a}\left( \omega \right) $ and $\widetilde{b}\left( \omega \right) 
$ is carried out in the Appendix. Further simplification is obtained under
the assumption that  the two quadrature noise components are totally
uncorrelated [$\overline{\widetilde{a}\left( \omega \right) \widetilde{b}%
\left( \omega ^{\prime }\right) }=\overline{\widetilde{a}\left( \omega
\right) \widetilde{b}^{\ast }\left( \omega ^{\prime }\right) }=0$]. The
expression for the cross-correlation spectrum $S_{qq^{\prime }}\left( \omega
\right) \equiv \overline{\widetilde{\delta I}_{q}\left( \omega \right) 
\widetilde{\delta I}_{q^{\prime }}^{\ast }\left( \omega \right) }$ is
obtained in the Appendix in the form: 
\begin{equation}
S_{qq^{\prime }}\left( \omega \right) =G_{qq^{\prime }}^{a}\left( \omega
\right) \overline{\left| \widetilde{a}\left( \omega \right) \right| ^{2}}%
+G_{qq^{\prime }}^{b}\left( \omega \right) \overline{\left| \widetilde{b}%
\left( \omega \right) \right| ^{2}}  \label{Sqq' nueva}
\end{equation}
(see Eq. \ref{espectro con ruidos de amplitud y fase}) showing that the
intensity noise spectrum can be derived from the knowledge of the noise
power spectra on the two quadratures of the incident field. It should be
mentioned that in the case of diode lasers, correlations between phase and
amplitude fluctuations are known to play a significant role \cite
{HENRY82,VAHALA83} and consequently Eq. \ref{Sqq' nueva} should only be
considered as approximative. However, in the following section the amplitude
fluctuations will be neglected ($\widetilde{a}\left( \omega \right) =0$)\ in
which case Eq. \ref{Sqq' nueva} can be consistently used. 

In a balanced detection experiment, two photodiodes record the light
intensities corresponding to two orthogonal polarizations $\mathbf{q}$ and $%
\mathbf{q}^{\prime }$ respectively. The fluctuation power spectra of the sum
[$Z_{+}\left( \omega \right) $] and difference [$Z_{-}\left( \omega \right) $%
] of the two photocurrents are given by: 
\begin{equation}
Z_{\pm }\left( \omega \right) =S_{qq}\left( \omega \right) +S_{q^{\prime
}q^{\prime }}\left( \omega \right) \pm S_{qq^{\prime }}\left( \omega \right)
\pm S_{q^{\prime }q}\left( \omega \right)
\label{espectros suma y diferencia}
\end{equation}

\section{Numerical results.}

We present in this section the transmitted field noise spectra calculated
from the model above. Having the comparison with experimental results in
mind, we restrict ourselves to a limited choice of the parameters within the
large freedom allowed by the model. We consider a model atomic transition $%
F_{g}=1\rightarrow F_{e}=0$. The transition is considered closed in the
sense that the radiative decay from the excited state returns the atom to
the ground level ($\eta =1$ in the model). However, the transition actually
behaves as an open transition due to the existence of a trapping state in
the ground level~\cite{RENZONI99}. Numerical simulations with more realistic
choices of the atomic level angular momenta and branching ratio are
straightforward and give results qualitatively similar to those presented
here.

The field incident on the atomic sample is taken linearly polarized and we
consider the decomposition of the transmitted light either into two
orthogonal circular polarization components or two orthogonal linear
polarization components.

A strong simplification is introduced based on the fact that phase noise is
known to be dominant in diode lasers \cite{GIACOBINO95}. We consider that
all the incident field noise is in the quadrature component $b\left(
t\right) $ [we take $a\left( t\right) =0$]. Such assumption leads to a
qualitative agreement with the experimental observation. If amplitude noise
is introduced in the calculation, it generally results in a dominant noise
background since in this case there are intensity fluctuations at the
detectors even in the absence of the atomic medium.

We calculate noise power at a fixed analysis frequency $\omega $ as a
function of the longitudinal magnetic field $B$. This calculation does not
require the knowledge of the incident field spectrum. According to Eq. \ref
{Sqq' nueva}, only the value of $\overline{\left| \widetilde{b}\left( \omega
\right) \right| ^{2}}$  at the analysis frequency (a scale factor) is
needed. In all the calculations presented below we have used $\omega
=0.3\Gamma $ ($\sim $ 2 MHz in the case of the Rb D lines). The evaluations
of $\beta _{q}$, and $\widetilde{\delta \beta }_{q}\left( \omega \right) $
require the knowledge of the proportionality constant $K$ (see Eqs. \ref
{betaq}, \ref{fb de omega}). We have taken $K=1$ which (together with $%
\gamma =10^{-3}\Gamma $) results in $\sim $20\% weak field linear absorption
(a rather realistic figure in view of the experiments).

\begin{figure}[tbp]
\includegraphics[width=3.5in]{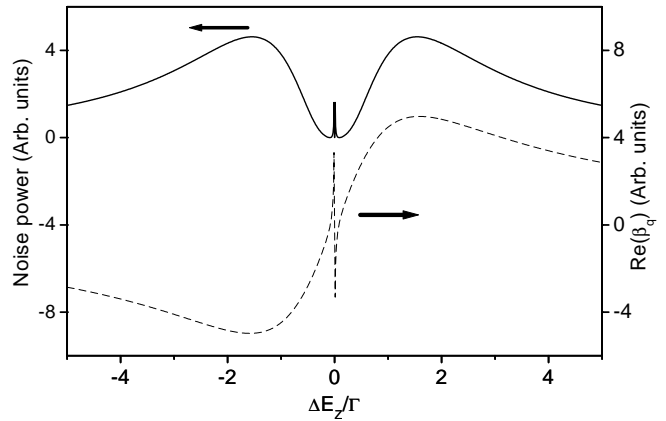}
\caption{Solid line: noise power at $\protect\omega =0.3\Gamma $ of a
circular polarization component of the transmitted light field after
interaction with an $F_{g}=1\rightarrow F_{e}=0$ atomic transition. Dashed
line: real part of the mean atomic dipole along the considered circular
polarization. Parameters: $\protect\delta =0$, $\Omega =0.4\Gamma $, $%
\protect\gamma =10^{-3}\Gamma $, $K=1 $.}
\label{Noise cc broad}
\end{figure}

The calculated value for the noise power $S_{qq}\left( \omega \right) $ for
one circular polarization component is presented as a function of the
longitudinal magnetic field in Fig.~\ref{Noise cc broad} (in terms of the
corresponding Zeeman energy shift: $\Delta E_{Z}=g\mu _{B}B$). In this plot,
the atom field detuning $\delta =\omega _{0}-\omega _{L}$ is taken as zero.
The noise spectrum is mainly composed of two broad peaks symmetric around $%
B=0$ with width and separation of the order of $\Gamma $ and a narrow
structure (a doublet) of characteristic width much smaller than $\Gamma $.
The broad structure arises from a linear magneto-optical effect: the
circular birefringence resulting in the dephasing of one circular field
component with respect to the incident field. As discussed above, such
dephasing results in phase to amplitude noise conversion. The central
structure is a result of the NMOE occurring near zero magnetic field owing
to ground state coherence. For comparison, we show in the same figure the
real part of the mean induced atomic dipole along the circular polarization
considered. Notice the steep variation around $B=0$. The zero-crossing of
the   real part of the mean induced atomic dipole results in the doublet
structure and the exact cancellation of the noise at $B=0$. At low
intensities the width of the narrow structure is determined by $\gamma $.

\begin{figure}[tbp]
\includegraphics[width=3.5in]{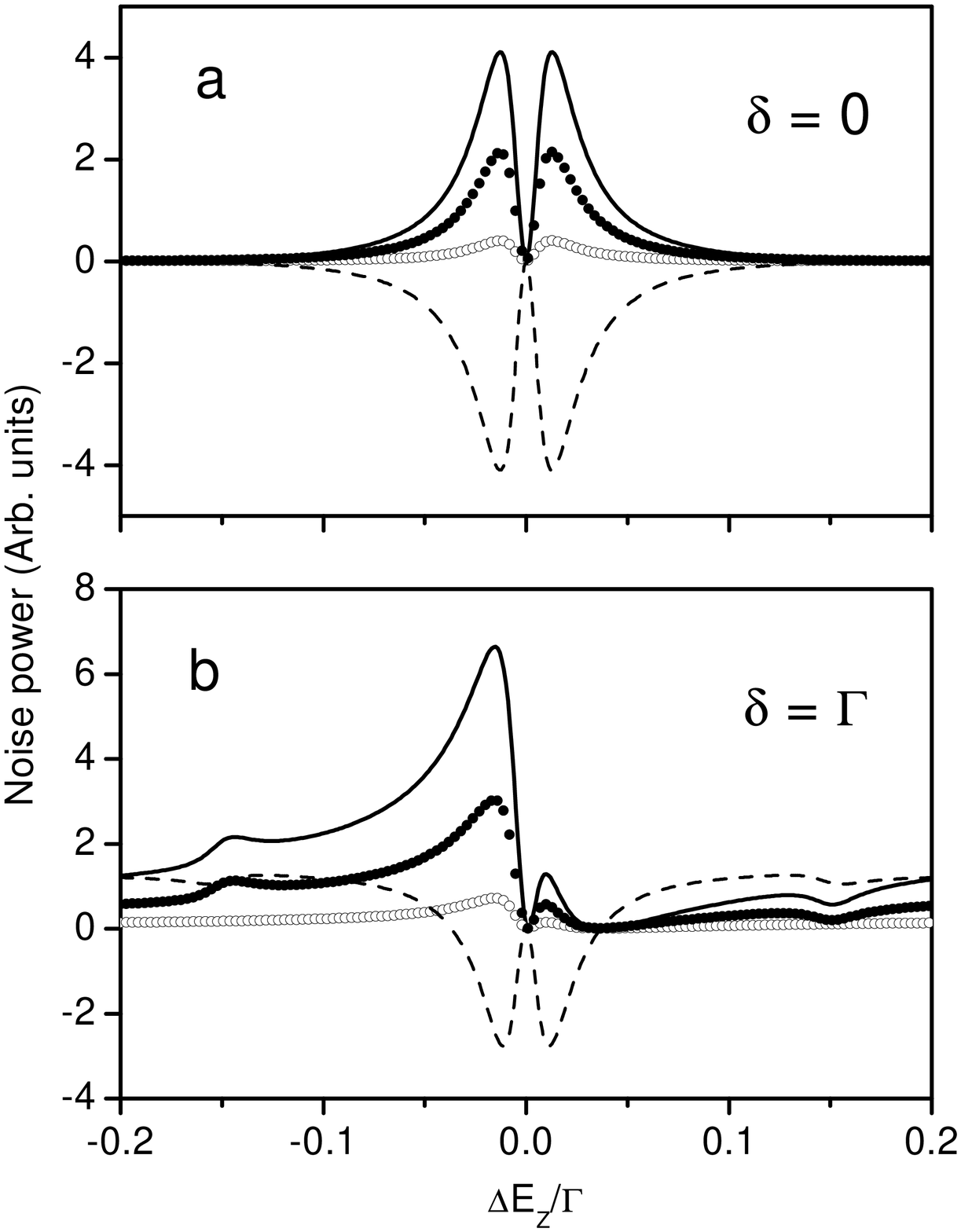}
\caption{Nonlinear magneto-optical effect on noise power. Solid line: noise
power $S_{qq}\left( \protect\omega \right) $ for a single circular
component. Solid circles: contribution to $S_{qq}\left( \protect\omega %
\right) $ of $\widetilde{\protect\delta I}_{q}^{A}\left( \protect\omega
\right) $. Hollow circles: contribution to $S_{qq}\left( \protect\omega %
\right) $ of $\widetilde{\protect\delta I}_{q}^{F}\left( \protect\omega
\right) $. Dashed line: cross-correlation noise power $S_{qq^{\prime
}}\left( \protect\omega \right) $ for the two orthogonal circular
components. a) $\protect\delta =0$, b) $\protect\delta =\Gamma $. Other
parameters: $\protect\omega =0.3\Gamma $, $\Omega =\Gamma $, $\protect\gamma %
=10^{-3}\Gamma $, $K=1$.}
\label{Noise cc central}
\end{figure}

A closer look into the central NMO structure is presented in Fig. \ref{Noise
cc central}. The plot of the noise power $S_{qq}\left( \omega \right) $
(solid line) is shown together with the contributions to the noise power of
the two terms $\widetilde{\delta I}_{q}^{F}\left( \omega \right) $ and $%
\widetilde{\delta I}_{q}^{A}\left( \omega \right) $ identified in Eq. \ref
{delta I de omega} (circles). Also shown in Fig. \ref{Noise cc central}a is
the cross-correlation spectrum $S_{qq^{\prime }}\left( \omega \right) $ for
the two different (orthogonal) circular polarizations. In the present case\
of $\delta =0$, we have $S_{qq^{\prime }}\left( \omega \right)
=-S_{qq}\left( \omega \right) $ and consequently $Z_{+}\left( \omega \right)
=0$ and $Z_{-}\left( \omega \right) =4S_{qq}\left( \omega \right) $.

In figure \ref{Noise cc central}b a nonzero laser-atom detuning was used. We
have taken $\delta =\Gamma $ as a representative example. Notice the
asymmetry in the plot of $S_{qq}\left( \omega \right) $. Such asymmetry is
expected since, for zero detuning, a given circular polarization of the
field will get closer or farther to resonance depending on the magnetic
field sign. Nevertheless, the cross-correlation spectrum $S_{qq^{\prime
}}\left( \omega \right) $ remains symmetric with respect to $B=0$. Another
significant feature in Fig. \ref{Noise cc central}b is the presence in both $%
S_{qq}\left( \omega \right) $ and $S_{qq^{\prime }}\left( \omega \right) $
of sidebands occurring for values of the magnetic field for which the Zeeman
shift $\Delta E_{Z}$ satisfies $2\Delta E_{Z}=\hbar \omega $ ($\Delta
E_{Z}/\Gamma =0.15$ for the parameters of Fig.~\ref{Noise cc central})~\cite
{BUDKER02B}. One can check from Fig. \ref{Noise cc central}b that these
sidebands are entirely due to the $\widetilde{\delta I}_{q}^{A}\left( \omega
\right) $ contribution.

\begin{figure}[tbp]
\includegraphics[width=3.5in]{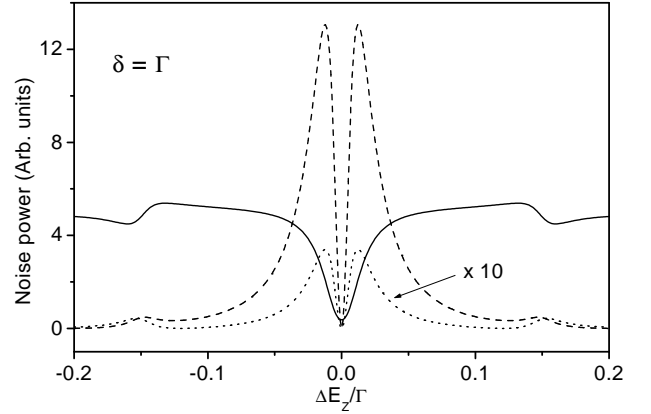}
\caption{Calculated balanced detector noise signals. Dashed: photocurrent
difference noise for circular polarizations. Dotted: photocurrent difference
noise for linear polarizations. Solid: photocurrent sum noise for both
polarization choices. $\protect\delta =\Gamma $, $\protect\omega =0.3\Gamma $%
, $\Omega =\Gamma $, $\protect\gamma =10^{-3}\Gamma $, $K=1$.}
\label{Sum dif}
\end{figure}

The plots of the calculated intensity sum and difference noise signals, $%
Z_{+}\left( \omega \right) $ and $Z_{-}\left( \omega \right) $ respectively,
for the choice of parameters of Fig.~\ref{Noise cc central}b in a balanced
polarimeter in which the field is decomposed into orthogonal circular
polarization components are shown in Fig. \ref{Sum dif}. \ $Z_{+}\left(
\omega \right) $, the total transmitted intensity noise, presents a broad
background level owing to \emph{linear} phase to amplitude noise conversion
produced by the nonzero real part of the mean atomic polarization. Along
with this broad linear contribution, $Z_{+}\left( \omega \right) $ presents
narrow NMO structures: the central dip and the sidebands. The central dip is
easily understood by noticing that at $B=0$ the condition is met for
coherence population trapping (CPT) among Zeeman sublevels which result in
electromagnetically induced transparency (EIT). The sidebands are due to
resonances in the atomic fluctuations when the analysis frequency
corresponds to twice the atomic Larmor frequency. The photocurrent
difference noise $Z_{-}\left( \omega \right) $ is dominated by the central
doublet which is related to the nonlinear circular birefringence occurring
around $B=0$. For a nonzero magnetic field, the two circular polarizations
of the induced atomic dipole are dephased in opposite directions resulting
in anti-correlated light fluctuations. As a result of the symmetry existing
at $B=0$ for the two polarization components we have $S_{qq}\left( \omega
\right) =S_{q^{\prime }q^{\prime }}\left( \omega \right) =S_{qq^{\prime
}}\left( \omega \right) =S_{q^{\prime }q}\left( \omega \right) $ and
consequently $Z_{-}\left( \omega \right) $ cancels at this point. The
dependence of $\ Z_{-}\left( \omega \right) $ (for orthogonal circular
polarizations) on the light intensity is presented in Fig. \ref{dep
intensidad teo}.

\begin{figure}[tbp]
\includegraphics[width=3.5in]{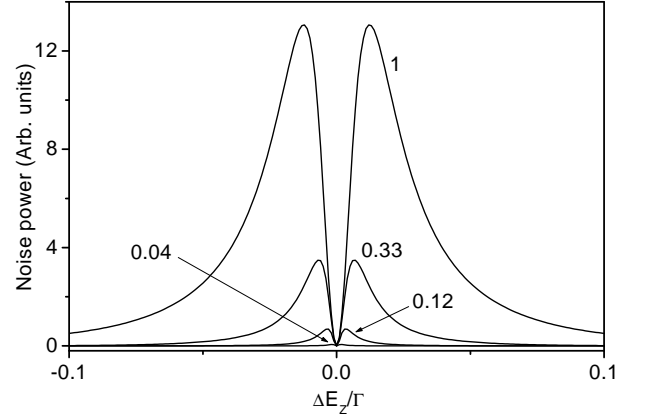}
\caption{{}Intensity dependence of the balanced detector photocurrent
difference noise signal $Z_{-}\left( \protect\omega \right) $ for circular
polarizations. The corresponding values of $\left( \Omega /\Gamma \right)
^{2}$ are indicated ($\protect\delta =\Gamma $, $\protect\omega =0.3\Gamma $%
, $\protect\gamma =10^{-3}\Gamma $, $K=1$).}
\label{dep intensidad teo}
\end{figure}

\begin{figure}[tbp]
\includegraphics[width=3.5in]{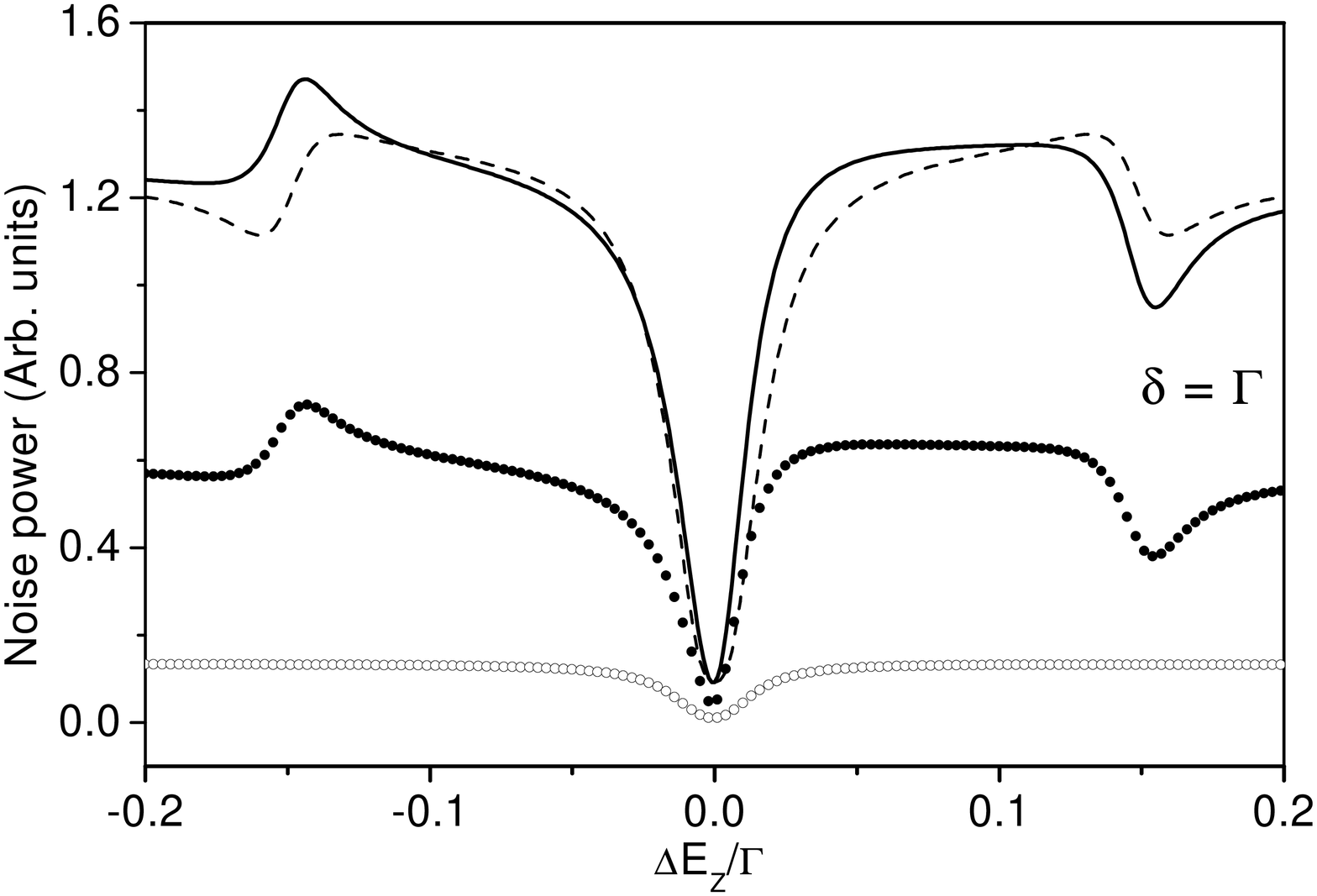}
\caption{Nonlinear magneto-optical effect on noise power. Solid line: noise
power $S_{qq}\left( \protect\omega \right) $ for a single linear
polarization component. Solid circles: contribution to $S_{qq}\left( \protect%
\omega \right) $ of $\widetilde{\protect\delta I}_{q}^{A}\left( \protect%
\omega \right) $. Hollow circles: contribution to $S_{qq}\left( \protect%
\omega \right) $ of $\widetilde{\protect\delta I}_{q}^{F}\left( \protect%
\omega \right) $. Dashed line: cross-correlation noise power $S_{qq^{\prime
}}\left( \protect\omega \right) $ for the two orthogonal linear polarization
components. $\protect\delta =\Gamma $, $\protect\omega =0.3\Gamma $, $\Omega
=\Gamma $, $\protect\gamma =10^{-3}\Gamma $, $K=1$.}
\label{Noise ll central}
\end{figure}

Let us discuss now the case of a balanced detector polarimeter in which the
two orthogonal output polarizations $\mathbf{q}$ and $\mathbf{q}^{\prime }$
are linear. In such a case, if the laser-atom detuning $\delta $ is zero the
output light fluctuations are zero on both polarizations. This is a somehow
surprising result since the nonlinear Faraday effect results in the rotation
of the mean transmitted field polarization. However, in this case the real
part of the induced atomic polarization along the two components $\mathbf{q}$
and $\mathbf{q}^{\prime }$ is zero (i.e. the interaction is purely
absorptive) and thus no phase to noise conversion occurs. The situation is
different for nonzero $\delta $. The curves corresponding to $S_{qq}\left(
\omega \right) $ and $S_{qq^{\prime }}\left( \omega \right) $ for $\delta
=\Gamma $ are presented in Fig. \ref{Noise ll central}. The corresponding
trace for $Z_{-}\left( \omega \right) $ is shown in Fig. \ref{Sum dif} $\ $($%
Z_{+}\left( \omega \right) $ is the same than for circular polarizations).
Notice the significant reduction of $Z_{-}\left( \omega \right) $ in
comparison with the result obtained for circular polarizations. Such
behavior is in contrast with the experimental observation where an increase
in the photocurrent difference noise is observed for linear polarization. We
will discuss this discrepancy in the next section.

\section{Discussion.}

The traces presented in Fig. \ref{Signal} provide a clear demonstration of
the use of noise analysis for the observation of NMOE. The observed features
(much narrower than the optical transition linewidth $\Gamma $) are the
consequence of ground state coherence. As in the usual ground state Hanle
effect and other NMOE, a coherence resonance is observed around $B=0$.
Additional resonances of comparable width are also observed when the
magnetic field induces a precession of the ground state alignment at one
half of the analysis frequency. Such resonances may be understood as Raman
resonances between the laser carrier frequency and the ``sidebands''
produced by the light ``modulation'' produced by the spectral component of
the noise at the analysis frequency.

The similarity between the experimental observations and the calculations
(Fig. \ref{Sum dif}) is apparent. It suggests that the essential ingredients
of the NMO noise effects are contained in the simple model considered.

However, several important differences between the model assumptions and the
actual experimental conditions must be mentioned. The first concerns the
fact that the numerical calculations were made for a homogeneous sample of
atoms while the experiment is carried out on a vapor cell where different
atoms experience different detunings $\delta $, owing to the Doppler effect.
Numerical integration over $\delta $ is straightforward. The resulting
curves for the balanced detection signals preserve the shape of the central
feature but do not present the relatively large weight of the sidebands
observed in the experiments. Our second comment concerns the naive
description of the laser noise. As already mentioned, the laser amplitude
noise was completely neglected. In the context of the model this appears to
be a reasonable assumption since even modest amounts of amplitude noise
(compared to quadrature noise) result in a magnetic-field-independent signal
background. Moreover, diode lasers are known to possess phase noise in
excess of amplitude noise over a wide bandwidth. Nevertheless, this
simplification is quite restrictive since in addition to ignoring the small
but non-negligible amplitude noise present in diode lasers, it prevents the
description within the model of the effect of shot-noise. Another important
simplification in the theory above is the linearized treatment of the field
noise as due to small fluctuations around a constant mean value. Such a
treatment is not appropriate when the field phase experiences random
fluctuations resulting in phase diffusion. In the present experimental
conditions, the phase diffusion time of our laser is of the order of a
microsecond (inverse of the laser linewidth), which is short compared to the
temporal evolution of coherence resonances. This suggests that the treatment
of the laser light as a phase-diffusing field should be more appropriate. We
have carried out a calculation using a phase-diffusing field model that will
be presented elsewhere. 

A striking difference between the experimental and the theoretical curves
concerns the relative amplitude of the photocurrent difference noise for
circular and linear polarizations. While the experimental result is that the
noise is enhanced for linear polarizations by a factor of approximately two,
the calculation shows a strong noise reduction (see Fig. \ref{Sum dif}). We
believe this to be a consequence of the inadequacy of the linear model. As a
matter of fact, preliminary results obtained with a phase diffusing field
model show, in this respect, a good agreement with the experimental
observation.

Another important difference between theoretical and experimental traces is
the weight of the sidebands. The large contribution to the noise occurring
at the sidebands appears to be out of reach for our present theoretical
approach. This suggests that new key ingredients such as correlated
amplitude and phase fluctuations, propagation effects in optically thick
media or the influence of resonance fluorescence scattered into the detected
light modes \cite{RIES03}, must be considered for a more accurate
theoretical description.

\section{Conclusions.}

We have shown that noise analysis provides a useful handle for the study of
NMOE. Our experiments reveal NMO features observed with large contrast and
resolution without the use of any light frequency modulation other than
noise modulation intrinsically present in any light source. The technique
presented here can be easily applied to other elements and transitions.
Since laser sources frequently present a broad noise spectrum, the noise
detection of NMOE can be used over a large dynamic range of analysis
frequencies, allowing the observation of coherent features for relatively
large values of the magnetic field (several Gauss). On the other hand,
application of the noise detection technique to paraffin coated or buffer
gas cells should allow the observation of coherence resonances in the $\mu $%
G range. Noise detected NMO resonances may therefore be of interest for high
precision magnetometry.

We have presented a simple and rather general theoretical treatment for the
calculation of the spectral properties of the noise signals available in a
balanced detection experiment. Good qualitative correspondence between the
predictions of the model and the observations was obtained. However, some
features, mainly the weight and shapes of the sideband resonances and the
observed dependence of the noise signals on the orthogonal polarization
decomposition, strongly suggest that a more precise treatment of the optical
field fluctuations (including quantum fluctuations) and the atomic medium
may be necessary. Additional work along this line is currently underway.

\begin{acknowledgments}
This work was supported by CSIC, PEDECIBA and Fondo Clemente Estable
(Uruguayan agencies) and FAPESP, CAPES and CNPq (Brazilian agencies).
\end{acknowledgments}

\appendix

\section{}

We present in this appendix the steps leading to the calculation of the
noise spectral signals observed in a balanced detection polarimeter. We
initially transform the Bloch equation Eq. \ref{bloch} using a Liouville
space representation. Later on, the linear response of the atomic dipole to
the incident fluctuating field is derived and the spectral signals
calculated.

\subsection{Liouville space representation.}

We consider a complex vectorial space $\mathbb{L}$ of dimension given by the
number of elements in $\rho $. To each density matrix $\rho $ in Hilbert
space we associate a (column) vector $y$ in $\mathbb{L}$ in such way that each
matrix element in $\rho $ corresponds to an element of $y$. 
\begin{equation}
\rho \longleftrightarrow y  \label{transformacion de Liouville}
\end{equation}
For an operator $O$ acting upon $\rho $ we have: 
\begin{eqnarray}
O\rho &\longleftrightarrow &\mathcal{L}\left( O\right) y
\label{operadores izquierda y derecha} \\
\rho O &\longleftrightarrow &\mathcal{R}\left( O\right) y \;,  \nonumber
\end{eqnarray}
where $\mathcal{L}\left( O\right) $ and $\mathcal{R}\left( O\right) $ are
linear operators in $\mathbb{L}$.

Let $u$ be the vector in $\mathbb{L}$ with all elements corresponding to
populations equal to one and all elements corresponding to coherences equal
to zero. We note for further use that $trace\left( O\rho \right) =u^{\dagger
}\mathcal{L}\left( O\right) y$.

After some algebra, one can show that the Bloch equation Eq. \ref{bloch} can
be written in the form \cite{WALSER94}: 
\begin{equation}
\dot{y}=\exp \left( -i\mathcal{N}\omega _{L}t\right) \mathcal{A}\exp \left( i%
\mathcal{N}\omega _{L}t\right) y+\gamma y_{0}  \label{Bloch_y}
\end{equation}
with: 
\begin{eqnarray}
\mathcal{N} &=&\mathcal{L}\left( P_{e}\right) \mathcal{-R}\left( P_{e}\right)
\label{Operador N} \\
\mathcal{A} &\equiv &-\frac{i}{\hbar }\left[ \mathcal{L}\left(
H_{0}+V\right) -\mathcal{R}\left( H_{0}+V\right) \right]  \nonumber \\
&&-\frac{\Gamma }{2}\left[ \mathcal{L}\left( P_{e}\right) +\mathcal{R}\left(
P_{e}\right) \right] -\gamma \mathbb{I}  \nonumber \\
&&+\eta \Gamma \left( 2F_{e}+1\right) \sum_{q=-1,0,1}\mathcal{L}\left(
Q_{ge}^{q}\right) \mathcal{R}\left( Q_{eg}^{q}\right)
\label{Operadores varios en Liouville} \\
\rho _{0} &\longleftrightarrow &y_{0}  \label{y0}
\end{eqnarray}
where $\mathbb{I}$ is the identity operator in $\mathbb{L}$ and $V=V_{ge}+V_{eg}$
(see Eq. \ref{interaccion campo atomo}).

It is convenient to consider the Liouville space vector $x=\exp \left( i%
\mathcal{N}\omega _{L}t\right) y$ entirely composed by slowly varying
coefficients. Substitution in Eq. \ref{Bloch_y} gives: 
\begin{equation}
\dot{x}=\left( i\mathcal{N}\omega _{L}+\mathcal{A}\right) x+\gamma y_{0}\;.
\label{ecuacion de Bloch para x}
\end{equation}

After solving \ref{ecuacion de Bloch para x}, the slowly varying envelope of
the atomic polarization can be obtained through: 
\begin{equation}
\mathbf{P}\left( t\right) =u^{\dagger }\mathcal{L}\left( \mathbf{D}%
_{eg}\right) x\left( t\right) \;.  \label{La polarizacion atomica}
\end{equation}

\subsection{Fluctuating atomic response.}

We seek a solution of Eq. \ref{ecuacion de Bloch para x} in the form $%
x\left( t\right) =\bar{x}+\delta x\left( t\right) $. Linearizing Eq. \ref
{ecuacion de Bloch para x} we have the following system: 
\begin{eqnarray}
0 &=&\mathcal{B}\left( \bar{\alpha}\right) \bar{x}+\gamma y_{0}
\label{Linearizada orden 0} \\
\delta \dot{x}\left( t\right) &=&\mathcal{B}\left( \bar{\alpha}\right)
\delta x\left( t\right)  \nonumber \\
&&-\left[ \mathcal{D}^{-}\delta \alpha \left( t\right) +\mathcal{D}%
^{+}\delta \alpha ^{\ast }\left( t\right) \right] \bar{x}
\label{linearizada orden 1} \\
\mathcal{D}^{-} &\equiv &\frac{i}{\hbar }\left[ \mathcal{L}\left( \mathbf{e}.%
\mathbf{D}_{ge}\right) -\mathcal{R}\left( \mathbf{e}.\mathbf{D}_{ge}\right) %
\right]  \label{D-} \\
\mathcal{D}^{+} &\equiv &\frac{i}{\hbar }\left[ \mathcal{L}\left( \mathbf{e}%
^{\ast }.\mathbf{D}_{eg}\right) -\mathcal{R}\left( \mathbf{e}^{\ast }.%
\mathbf{D}_{eg}\right) \right] \;,  \label{D+}
\end{eqnarray}
where $\mathcal{B}\equiv i\mathcal{N}\omega _{L}+\mathcal{A}$ and we have
taken advantage of the fact that $\mathcal{A}$ is linear on the complex
field amplitudes $\alpha \left( t\right) $ and $\alpha ^{\ast }\left(
t\right) $ (see Eq. \ref{Operadores varios en Liouville}). Eq. \ref
{Linearizada orden 0} gives the mean value $\bar{x}$ from which the average
atomic dipole components $\beta _{q}$ can be obtained (using Eq. \ref{La
polarizacion atomica}) as: 
\begin{equation}
\beta _{q}=Ku^{\dagger }\mathcal{L}\left( \mathbf{q}^{\ast }\cdot \mathbf{D}%
_{eg}\right) \bar{x}\;,  \label{betaq}
\end{equation}
where $K$ is a proportionality constant.

Taking the Fourier transform of Eq. \ref{linearizada orden 1}: 
\begin{equation}
\widetilde{\delta x}\left( \omega \right) =\left[ i\omega \mathbb{I}+\mathcal{B}%
\left( \bar{\alpha}\right) \right] ^{-1}\left[ \mathcal{D}^{-}\widetilde{%
\delta \alpha }\left( \omega \right) +\mathcal{D}^{+}\widetilde{\delta
\alpha }^{\ast }\left( -\omega \right) \right] \bar{x}\;,
\label{transformada de Fourier de deltax}
\end{equation}
we notice that $\widetilde{\delta x}\left( \omega \right) $ presents a
resonant behavior for $\omega $ approaching the eigenfrequencies of the
evolution operator $\mathcal{B}\left( \bar{\alpha}\right) $.

We proceed by writing $\delta \alpha \left( t\right) =a\left( t\right)
+ib\left( t\right) $, where $a\left( t\right) $ and $b\left( t\right) $ are
real fluctuating functions with zero mean value. Using the corresponding
Fourier transforms, $\widetilde{a}\left( \omega \right) $ and $\widetilde{b}%
\left( \omega \right) $, in Eq. \ref{transformada de Fourier de deltax} we
have: 
\begin{eqnarray}
\widetilde{\delta x}\left( \omega \right) &=&\left[ i\omega \mathbb{I}+\mathcal{%
B}\left( \bar{\alpha}\right) \right] ^{-1}\left[ \left( \mathcal{D}^{-}+%
\mathcal{D}^{+}\right) \widetilde{a}\left( \omega \right) \right.  \nonumber
\\
&&\left. +i\left( \mathcal{D}^{-}-\mathcal{D}^{+}\right) \widetilde{b}\left(
\omega \right) \right] \bar{x}\;,  \label{delta x de omega}
\end{eqnarray}
and from Eq. \ref{La polarizacion atomica} we get: 
\begin{equation}
\widetilde{\delta \beta }_{q}\left( \omega \right) =f_{a}\left( \omega
\right) \widetilde{a}\left( \omega \right) +f_{b}\left( \omega \right) 
\widetilde{b}\left( \omega \right) \;,  \label{introduce fa y fb}
\end{equation}
with 
\begin{eqnarray}
f_{a}\left( \omega \right) &=&Ku^{\dagger }\mathcal{L}\left( \mathbf{q}%
^{\ast }\cdot \mathbf{D}_{eg}\right) \left[ i\omega \mathbb{I}+\mathcal{B}%
\left( \bar{\alpha}\right) \right] ^{-1}  \nonumber \\
&&\times \left( \mathcal{D}^{-}+\mathcal{D}^{+}\right) \bar{x}
\label{fa de omega} \\
f_{b}\left( \omega \right) &=&Ku^{\dagger }\mathcal{L}\left( \mathbf{q}%
^{\ast }\cdot \mathbf{D}_{eg}\right) \left[ i\omega \mathbb{I}+\mathcal{B}%
\left( \bar{\alpha}\right) \right] ^{-1}  \nonumber \\
&&\times i\left( \mathcal{D}^{-}-\mathcal{D}^{+}\right) \bar{x}\;.
\label{fb de omega}
\end{eqnarray}

Substituting Eq. \ref{introduce fa y fb} into Eqs. \ref{delta I de omega}-%
\ref{delta I atomo} we write: 
\begin{equation}
\widetilde{\delta I}_{q}\left( \omega \right) =g_{q}^{a}\left( \omega
\right) \widetilde{a}\left( \omega \right) +g_{q}^{b}\left( \omega \right) 
\widetilde{b}\left( \omega \right) \;,
\label{delta Iq en funcion de ga y gb}
\end{equation}
with 
\begin{eqnarray}
g_{q}^{a}\left( \omega \right) &\equiv &\left( \alpha _{q}^{\ast }-i\beta
_{q}^{\ast }\right) \left[ \left( \mathbf{q}^{\ast }\cdot \mathbf{e}\right)
+if_{a}\left( \omega \right) \right]  \nonumber \\
&&+\left( \alpha _{q}+i\beta _{q}\right) \left[ \left( \mathbf{q}\cdot 
\mathbf{e}^{\ast }\right) -if_{a}^{\ast }\left( -\omega \right) \right]
\label{ga} \\
g_{q}^{b}\left( \omega \right) &\equiv &i\left( \alpha _{q}^{\ast }-i\beta
_{q}^{\ast }\right) \left[ \left( \mathbf{q}^{\ast }\cdot \mathbf{e}\right)
+f_{b}\left( \omega \right) \right]  \nonumber \\
&&-\left( \alpha _{q}+i\beta _{q}\right) \left[ \left( \mathbf{q}\cdot 
\mathbf{e}^{\ast }\right) +f_{b}^{\ast }\left( -\omega \right) \right] \ .
\label{gb}
\end{eqnarray}

\subsection{Spectrum.}

We are interested in the spectral components: 
\begin{equation}
S_{qq^{\prime }}\left( \omega \right) =\overline{\widetilde{\delta I}%
_q\left( \omega \right) \widetilde{\delta I}_{q^{\prime }}^{*}\left( \omega
\right) } \;.  \label{espectro de la correlacion qq'}
\end{equation}
Further simplification can be obtained under the additional assumption that
the two quadrature components of the noise are uncorrelated [$\overline{%
\widetilde{a}\left( \omega \right) \widetilde{b}\left( \omega ^{\prime
}\right) }=\overline{\widetilde{a}\left( \omega \right) \widetilde{b}%
^{*}\left( \omega ^{\prime }\right) }=0$]; using Eq. \ref{delta Iq en
funcion de ga y gb} we get: 
\begin{eqnarray}
S_{qq^{\prime }}\left( \omega \right) &=&g_q^a\left( \omega \right)
g_{q^{\prime }}^{a*}\left( \omega \right) \overline{\left| \widetilde{a}%
\left( \omega \right) \right| ^2}  \nonumber \\
&&+g_q^b\left( \omega \right) g_{q^{\prime }}^{b*}\left( \omega \right) 
\overline{\left| \widetilde{b}\left( \omega \right) \right| ^2} \;.
\label{espectro con ruidos de amplitud y fase}
\end{eqnarray}

\end{document}